\documentclass[12pt]{iopart}
\usepackage[dvips]{graphics,color}
\usepackage{iopams,wrapfig}
\usepackage{xspace}
\usepackage{subfig}
\begin{document}

\title{STAR's Measurement 
of Energy and System Size Dependence of K/$\pi$ Fluctuations at RHIC.} 

\author{Zubayer Ahammed for the STAR Collaboration}

\address{ Variable Energy Cyclotron Centre,
1/AF, BidhanNagar, Kolkata-700064, INDIA}
\ead{za@veccal.ernet.in}
\begin{abstract}
Strangeness enhancement has been predicted to be one of the signatures of the formation of the quark gluon plasma (QGP) and  event-by-event
fluctuations of strangeness may reveal the  nature of such a phase
transition.
We report results for $K/\pi$ fluctuations from Cu+Cu at $\sqrt{s_{NN}}$ = 62.4
 and 200 GeV and Au+Au collisions at $\sqrt{s_{NN}}$ = 62.4, 130 and 200
GeV using the STAR detector at the Relativistic Heavy Ion Collider.
We compare our results 
with the  results  observed in central Pb+Pb collisions at SPS
energies at $\sqrt{s_{NN}}$ = 6.3, 7.6, 8.8, 12.3, and 17.3 GeV. 
We observe the fluctuation strength measured at the highest SPS energy point by the NA49 collaboration in Pb+Pb collisions is similar to our results  in Cu+Cu and Au+Au collisions. 
We also compare our results with HIJING, 
a statistical hadronization model and a multi-phase transport model. 
\end{abstract}


\section{Introduction}
Quantum Chromodynamics(QCD) predicts a transition of normal hadronic matter to
a new state of matter at sufficient high temperature or energy density. 
This new state of matter is known as Quark Gluon Plasma(QGP). 
Event-by-event fluctuations in global observables is considered to be one 
of the signatures of QGP phase transition. The nature of this phase 
transition may also be revealed in the fluctuations of global observables. 
Kapusta {\cite{meik}}et al, predicted that fluctuations in 
kaon to pion ratio may reveal the nature such a phase transition.
NA49 recently reported the measurement of strangeness production at
several beam energies lower than RHIC(200 GeV). A "horn" like discontinuity in $K^{+}/\pi^{-}$ 
ratio and continuity in $K^{-}/\pi^{-}$ ratio observed in Pb+Pb collisions has generated lot 
of excitement to study strangeness fluctuations event wise{\cite{horn}}. 
In the present talk, we report STAR's measurement of $K/\pi$ fluctuations on 
an event-by-event basis for Cu+Cu collision at 62.4 and 200 GeV and for Au+Au collision at 
62.4, 130 and 200 GeV.
\section{Measures}
Event-by-event fluctuations consist of mainly two components, the statistical
fluctuations arising due to finite number statistics of particle production and the dynamical component due to particle production mechanism. So its
necessary to remove the statistical fluctuations from the total fluctuations to get the 
dynamical component of fluctuations. We have used two measures to quantify the 
dynamical fluctuation strength. In first case, the statistical part is 
being estimated by constructing mixed events. The mixed events are being constructed taking one particle from each real event randomly keeping the multiplicity
of the mixed events same as real events. We use  a measure, $\sigma_{dyn,K\pi}$ as:
\begin{equation*}
\sigma_{dyn,K\pi}=\sqrt{\sigma_{data}^{2}-\sigma_{mixed}^{2}}
\end{equation*}
where $\sigma_{data}$ is the width of the $K/\pi$ distribution of real data and $\sigma_{mixed}$ that of mixed events. Then  $\sigma_{dyn,K\pi}$ is divided by Mean of $K/\pi$ distribution of real data and multiplied by 100 to get the percentage of fluctuation strength.
However the variable $\sigma_{dyn}$ may be problematic at lower multiplicities, so we  use another measure, $\nu_{dyn,K\pi}$
defined as :
\begin{equation*}
\nu_{dyn,K\pi}=\frac{<N_{K}(N_{K}-1)>}{<N_{K}>^2}+\frac{<N_{\pi}(N_{\pi}-1)>}{<N_{\pi}>^2}-2\frac{<N_{K}N_{\pi}>}{<N_{K}><N_{\pi}>}
\end{equation*}
where, $N_{K}$ and $N_{\pi}$ are the number of kaons and pions in an event. The average is being done over large number of event sample being analyzed.
The variable $\nu_{dyn,K\pi}$ is assumed to be independent of tracking efficiencies
and was first used in STAR Collaboration for net charge 
fluctuation study{\cite{netc}}. 
\section{Particle Identification and track selection}
The data analyzed were measured using Time Projection Chamber(TPC) detector 
in STAR experiment located inside  solenoidal magnetic field. 
The particle identification is based on specific energy 
loss($dE/dx)${\cite{dedx}} measured in the TPC. In order to precisely identify a particle, we define  
$N{\sigma}_{X} = log(\frac{dE/dx}{B_{X}})/\sigma_{X}{\cite{nsigm}}$ for each particle 
where, X can be any particle type ($\pi^{\pm}$,$K^{\pm}$ etc.), $B_{X}$ is the expected mean $dE/dx$ of particle X, and $\sigma_{X}$ is the $dE/dx$ resolution of the TPC.
In the present analysis, all tracks within $-1<\eta<1$ and 
$200<pt<600$ MeV/c are selected.
We select a particle to be pion if $N{\sigma}_{\pi}<2$ and $N{\sigma}_{K}> 2$, similarly for kaon is selected if 
$N{\sigma}_{K}<2$ and $N{\sigma}_{\pi} > 2$.
For removing electron contamination, we give a tighter cuts on electron. 
A particle is called electron, if $N{\sigma}_{e}<1$. 
\section{Results and discussion}
We have shown the $K/\pi$ distribution for Cu+Cu collision at 200 GeV in 
Figure 1 (a). 
The filled circle in the figure corresponds to real data and the solid line corresponds to mixed events. 
\begin{figure}
\begin{center}
\hspace{0.0cm}
\rotatebox{0}{\resizebox{15.cm}{!}{\includegraphics{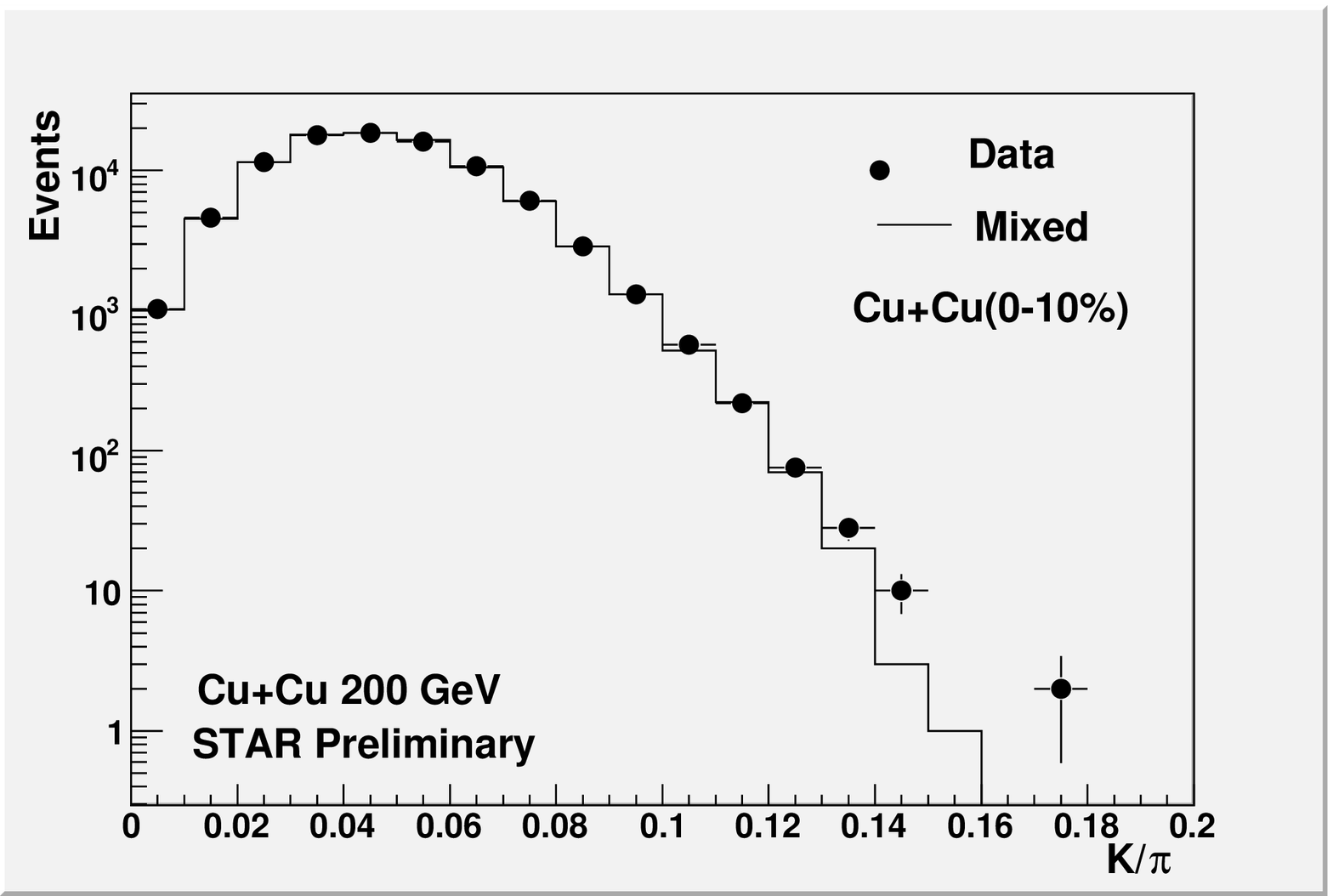}\includegraphics{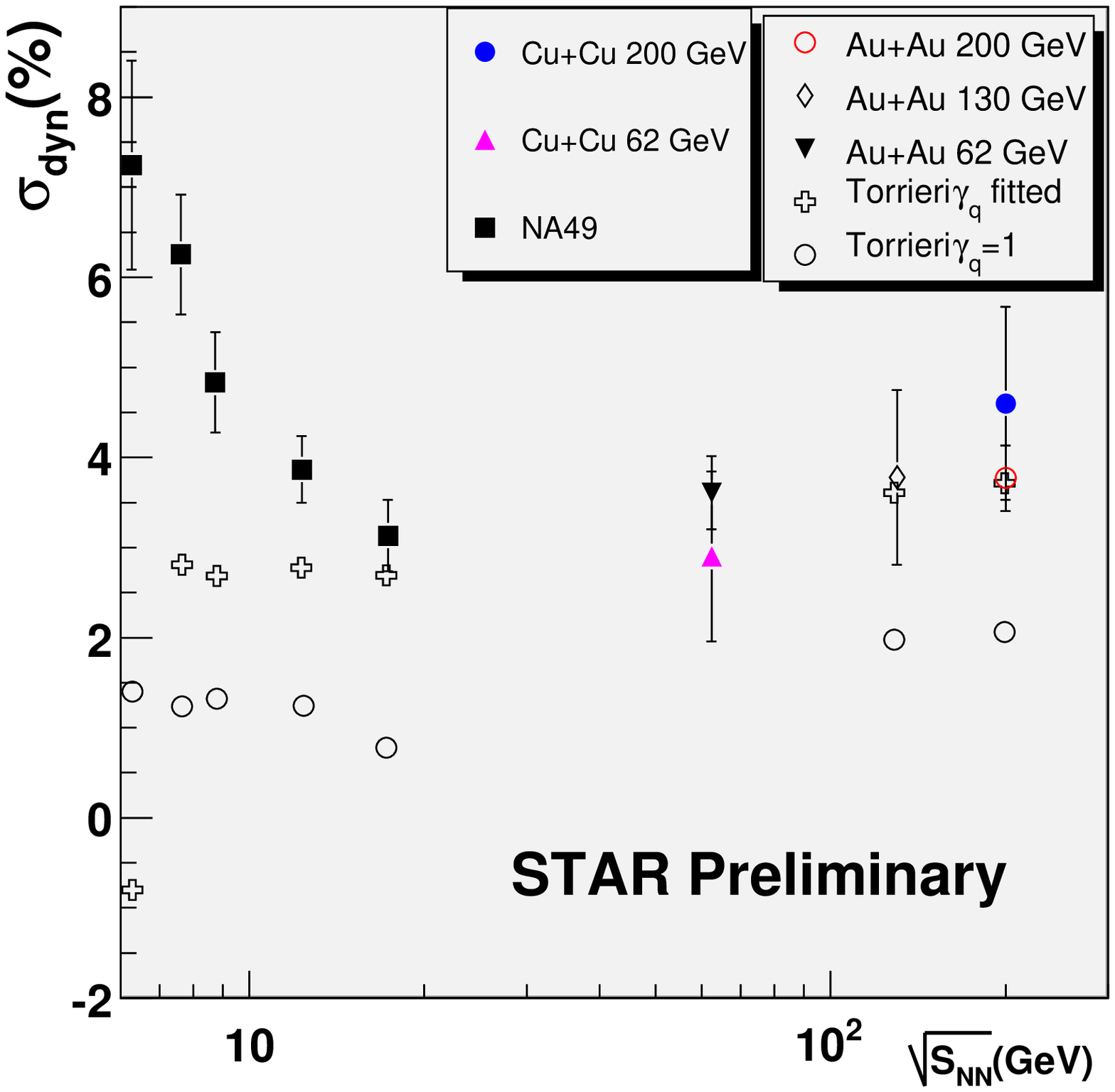}}}
\textbf{}
\begin{tabular*}{15.cm}{  l@{\extracolsep{7.5cm}}  l l @{\extracolsep{\fill}}}
\hspace{0.7cm} \textbf{(a)}  & \textbf{(b)}\\
\end{tabular*}
\caption{The $K/\pi$ ratio distribution for (a)Cu+Cu 200 GeV. The filled circles are from the data and the solid histogram corresponds to mixed events.The errors shown here are statistical only.
(b)The measured $\sigma_{dyn,K\pi}$ for Cu+Cu and Au+Au collision. The errors shown here for Au+Au and Cu+Cu are both statistical and systematic. Also shown in the figure is the comparison with the measurement of NA49 Collaboration.}\label{Figure1}
\vspace{-0.9cm}
\end{center}
\end{figure}
Using the width of this $K/\pi$ distributions,
$\sigma_{dyn,K\pi}$ is calculated. The measured $\sigma_{dyn,K\pi}$ is shown
in figure 1(b) for Cu+Cu 62.4 GeV and Cu+Cu 200 GeV. Also shown in the figure is the
\begin{figure}
\begin{center}
\hspace{0.0cm}
\rotatebox{0}{\resizebox{15.cm}{!}{\includegraphics{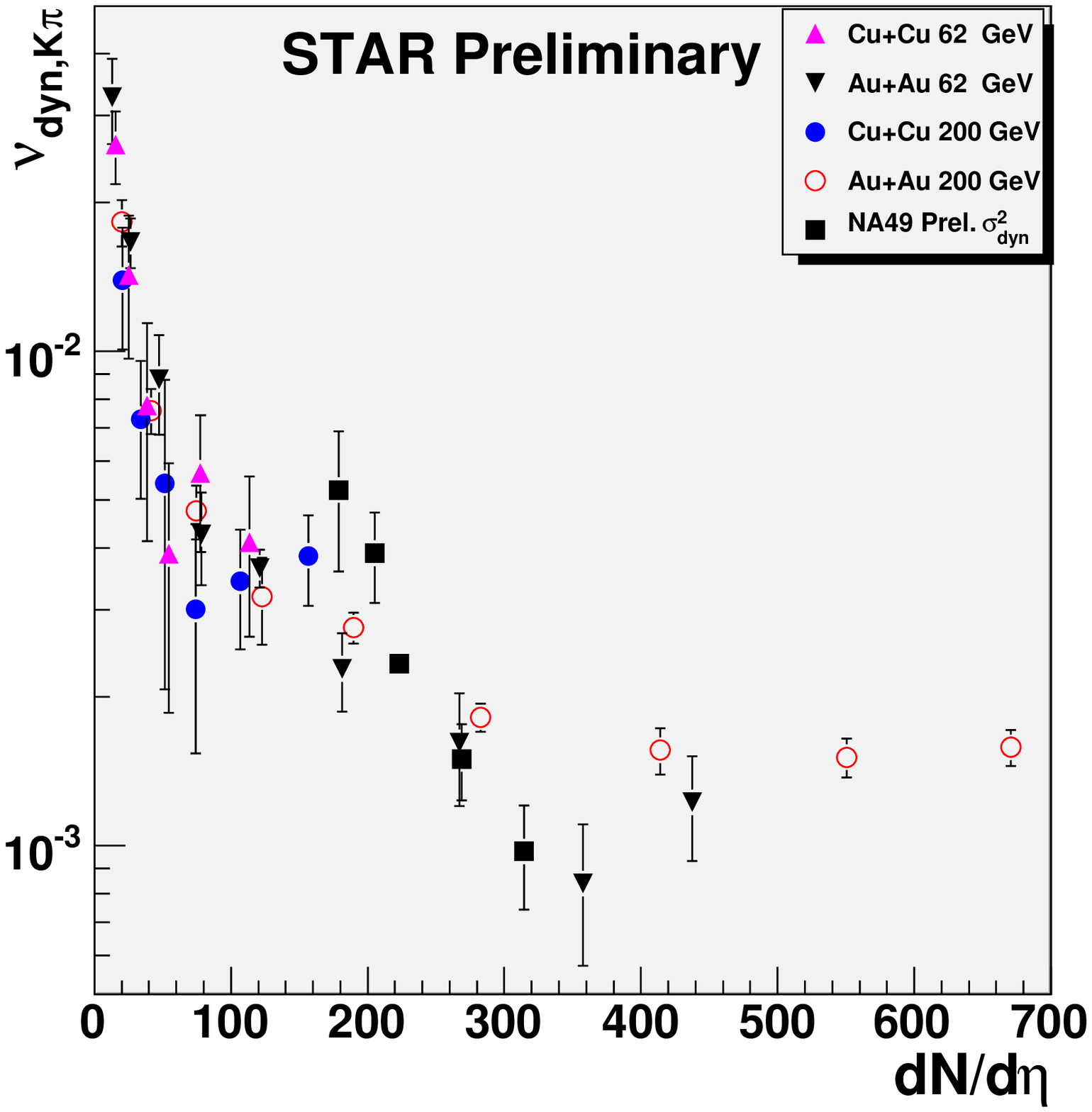}\includegraphics{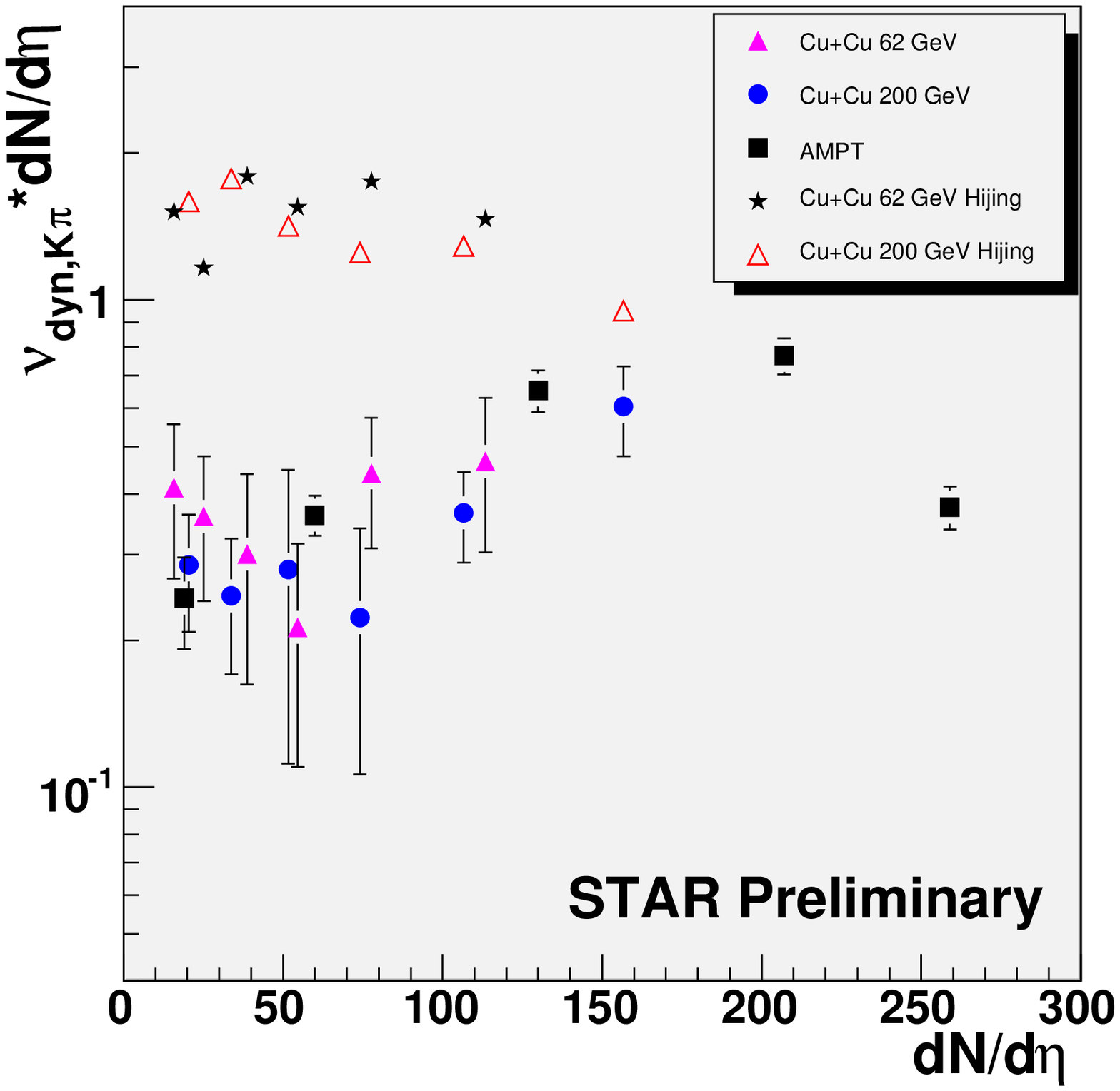}}}
\textbf{}
\begin{tabular*}{15.0cm}{ @{\extracolsep{7.5cm}} l l @{\extracolsep{\fill}}}
\hspace{-0.5cm} \textbf{(a)} & \textbf{(b)} \\
\end{tabular*}
\caption{(a)The measured $\nu_{dyn,K\pi}$ for Cu+Cu, Au+Au collisions.The errors shown for Cu+Cu collision includes statistical and systematic. However for Au+Au, only statistical errors shown. Systematic error for Au+Au is about 17.6$\%$.(b) The scaled $\nu_{dyn,K\pi}$ is compared to Corresponding results from HIJING and AMPT model calculations.  }\label{Figure2}
\end{center}
\end{figure}
$\sigma_{dyn,K\pi}$ for Au+Au collisions at 62.4, 130 and 200 GeV. We observe
the measured dynamical fluctuation 
strength is  independent of collision energy. It is seen that the $\sigma_{dyn,K\pi}$  is similar for Cu+Cu and 
Au+Au collisions. We have compared our results with the measurement of 
NA49 collaboration{\cite{cbl}}. The fluctuations strength measured at the highest SPS energy by 
NA49 collaboration in Pb+Pb collisions is similar to our results 
in Cu+Cu and Au+Au collisions within the present level of precision of 
measurement. The figure 1(b) also includes the predictions of 
Statistical Hadronization(SH) model{\cite{torr}}. 
The SH model predictions for non-equilibrium 
scenario
are consistent with our measurement at higher energy points, 
however they under predict the data at lower energies. In case of equilibrium scenario 
with fitted parameter,$\gamma_{q}=1$, the SH model under predicts our 
results at all energies.
The figure 2(a). shows the measured $\nu_{dyn,K\pi}$ for Cu+Cu and Au+Au collisions. 
Figure 2(a) also shows the comparison with NA49 measurement by using the relation 
${\sigma_{dyn,K\pi}}={\sqrt{\nu_{dyn,K\pi}}}$. Its observed that the measured $\nu_{dyn,K\pi}$ 
is similar to Cu+Cu and Au+Au collisions and independent of collision energy. 
We have also compared our results with model predictions mainly for HIJING model
and A Multi Phase Transport model(AMPT){\cite{ampt}}. AMPT model uses HIJING as initial particle
production in addition to hadronic evolution(multiple re-scattering). The track selection in HIJING and AMPT were done within same kinematic cuts as real data. The comparison of experimental results and the HIJING and AMPT predictions have been shown
in the figure 2(b). we observe that the HIJING over predicts the experimental 
results, however the AMPT predictions are in better agreement with data. 

\section{Summary}
We have measured $K/\pi$  fluctuation for Cu+Cu collisions at  62.4 and 200 GeV 
and for Au+Au collisions at 62.4,130 and 200 GeV. The measured fluctuation strength 
is observed to be independent of system size and energy. Our preliminary results 
are similar in comparison to top SPS energy measurements. HIJING model over predicts the 
experimental results however AMPT model is in better agreement with the 
data. 

\end{document}